\documentclass[aps,twocolumn,showpacs,preprintnumbers,amsmath,amssymb,footinbib]{revtex4}
\usepackage{mathtools}

\DeclarePairedDelimiter{\floor}{\lfloor}{\rfloor}
\usepackage{braket}
\usepackage{graphicx,epsfig}
\usepackage{bm}
\usepackage{dcolumn}
\usepackage{xcolor}
\usepackage[breaklinks=true,colorlinks,citecolor=blue,linkcolor=blue,urlcolor=blue]{hyperref}

\def\prv#1#2#3{Phys. Rev. {\bf #1}, #2 (#3)}
\def\rmp#1#2#3{Rev. Mod. Phys. {\bf #1}, #2 (#3)}
\def\prl#1#2#3{Phys. Rev. Lett. {\bf #1}, #2 (#3)}
\def\pra#1#2#3{Phys. Rev. A {\bf #1}, #2 (#3)}
\def\prb#1#2#3{Phys. Rev. B {\bf #1}, #2 (#3)}

\def\rpp#1#2#3{Rep. Prog. Phys. {\bf #1}, #2 (#3)}

\def\jpamt#1#2#3{J. Phys. A: Math. Theor. {\bf #1}, #2 (#3)}

\def\jpb#1#2#3{J. Phys. B: At. Mol. Opt. Phys. {\bf #1}, #2 (#3)}

\def\ajp#1#2#3{Am. J. Phys. {\bf #1}, #2 (#3)}

\def\jcp#1#2#3{J. Chem. Phys. {\bf #1}, #2 (#3)}

\def\noi{\noindent}
\def\bc{\begin{center}}
\def\ec{\end{center}}
\newcommand{\bea}{\begin{equation}}
\newcommand{\eea}{\end{equation}\noi}
\newcommand{\ber}{\begin{eqnarray}}
\newcommand{\eer}{\end{eqnarray}\noi}

\begin{document}
\title{Generalization of the Einstein coefficients and rate equations under the quantum Rabi oscillation}
\author{Najirul Islam}
\author{Shyamal Biswas}\email{sbsp [at] uohyd.ac.in}
\affiliation{School of Physics, University of Hyderabad, C.R. Rao Road, Gachibowli, Hyderabad-500046, India}


\date{February 07, 2022}

\begin{abstract}
We have generalized Einstein coefficients and rate equations from quantum field theoretic point of view by bringing the fundamental processes and the quantum Rabi oscillation in a single footing for the light-matter interactions for nonzero Rabi frequency. We have analytically obtained multimode Jaynes-Cummings model results for the quantum Rabi oscillations of a two-level system in a lossy resonant cavity containing (i) thermal photons and (ii) injected photons of a coherent field. We have renormalized the coupling constant for the light-matter interactions for these cases. Our results match well with the seminal experimental data obtained in this regard by Brune \textit{et al} [\href{http://dx.doi.org/10.1103/PhysRevLett.76.1800}{\prl{76}{1800}{1996}}]. We also have studied the population dynamics in this regard by applying the generalized Einstein rate equations.
\end{abstract}


\maketitle
 

\section{Introduction}
Seminal experimental work of Brune \textit{at al} \cite{Brune} regarding the quantum Rabi oscillation (or flopping) of the occupation of the two energy eigenstates of $^{87}$Rb atom in a lossy resonant cavity at finite temperatures, opened the possibilities of experimental \cite{Haroche-1998,Raimond,Miller} and theoretical \cite{Walther} study of the cavity quantum electrodynamics (QED) specially in the field of measuring and manipulation of individual quantum systems a quarter century back \cite{Brune,Meekhof,Wineland,Haroche}. The most interesting feature of the cavity-QED is that, the spontaneous emission from excited atoms or molecules can be greatly suppressed and enhanced by placing them in mirrors or in cavities, such as Fabry-Perot cavity, by virtue of the Purcell effect \cite{Purcell,Haroche-1998}. Experimentalists basically engineer the vacuum inside the cavity to observe the Purcell effect \cite{Haroche-1998}. Mode quality factor ($Q$) of resonant cavity plays an important role in this regard.  

A two-level system (atom or molecule) in the free space once makes a spontaneous emission, say at time $t=0$, the emitted photon goes away from the system in an irreversible manner. The possibility that after some finite time-interval the emitted photon would be further absorbed by the two-level system, was not considered in Einstein's semiclassical description \cite{Einstein}. However, observation of the quantum (vacuum) Rabi oscillation \cite{Brune} in the high-$Q$ cavity reveals the fact that, the boundary conditions greatly influence the atomic radiation \cite{Drexhage}, and consequently, the emitted photon is reabsorbed by the two-level system \cite{Haroche-1998,Raimond}. The spontaneous emission becomes reversible in an ideal\footnote{The mode quality factor goes to infinity for an ideal cavity.} cavity as the two-level system and the field exchange excitation at the rate of Rabi frequency ($\Omega_R$) \cite{Haroche-1998}. The periodicity in the exchange of the excitation leads to the time dependence in the Einstein coefficients.  

Three-dimensional multimode Jaynes-Cummings (J-C) Hamiltonian $\hat{H}=\frac{1}{2}\hbar\omega_0\sigma_3+\sum_{\vec{k}s}\hbar\omega_{\vec{k}}\hat{a}_{\vec{k}s}^\dagger\hat{a}_{\vec{k}s}-i\sum_{\vec{k}s}\hbar g_{\vec{k}s}[\sigma_+\hat{a}_{\vec{k}s}-\sigma_-\hat{a}_{\vec{k}s}^\dagger]$\footnote{Here we are following the notation \cite{Lahiri}: $\sigma_+=\ket{\psi_2}\bra{\psi_1}$, $\sigma_-=\ket{\psi_1}\bra{\psi_2}$, $\sigma_1=[\sigma_++\sigma_-]$, $\sigma_2=-i[\sigma_+-\sigma_-]$, $\sigma_3=\ket{\psi_2}\bra{\psi_2}-\ket{\psi_1}\bra{\psi_1}$, $\hat{a}_{\vec{k}s}$ ($\hat{a}_{\vec{k}s}^\dagger$) annihilates (creates) a photon of energy $\hbar\omega_{\vec{k}}$, polarization $s$ and momentum $\hbar\vec{k}$ (having dispersion $\omega=ck$) in the Fock space, $\ket{\psi_1}$ ($\ket{\psi_2}$) is the energy eigenstate for the lower (higher) energy $E_1$ ($E_2$) of the two-level system in absence of the light-matter interactions, $g_{\vec{k}s}$ is the coupling constant (assumed real) for the light-matter interaction for the mode $\vec{k}s$, and $\omega_0=(E_2-E_1)/\hbar$ is the Borh (angular) frequency of the two-level system.} \cite{Jaynes,Seke,Lahiri} which was proposed several decades back in this regard, is able to describe the quantum theory of radiation in a resonant cavity beyond (i) Dirac's determination of the Einstein $B$ coefficient \cite{Dirac} within the 1st order time-dependent perturbation theory of quantum mechanics and (ii) Weisskopf-Wigner determination of the Einstein $A$ coefficient \cite{Weisskopf} within the 1st order time-dependent perturbation theory of quantum field theory (quantum electrodynamics). Perturbation theories, however, can not explain the Rabi oscillation of a two-level system. The quantum Rabi oscillation, on the other hand, is well understood for the J-C model \cite{Jaynes} even for a single mode \cite{Scully,He}. This model basically offers an understanding of the light-matter interactions in terms of the fundamental processes (spontaneous emission, stimulated emission and absorption) in the light of the cavity-QED. Though there have been an enormous amount of theoretical investigation in the field of the cavity-QED \cite{Scully,Agarwal2}, nobody has come up with a cavity-QED theory, except a few quantum master equation approaches with the J-C model (for only the resonant mode) and a phenomenological damping \cite{Wilczewski,Chough} for the quantum Rabi oscillation of a two-level system in a lossy resonant cavity. Loss of the electromagnetic energy from the lossy resonant cavity, however, takes place for the frequency broadening around the resonant mode. This broadening naturally brings multimodes into account. 

Now we are coming up with a cavity-QED theory within the J-C model for multimodes around the resonance for explaining the quantum Rabi oscillation in a lossy resonant cavity as observed by Brune \textit{et al} \cite{Brune}. Multimode J-C model \cite{Seke} has become quite popular not only for an extension of the single-mode J-C model but also for the multi-photon transitions \cite{Li}, the dynamics of entanglement \cite{Shen}, \textit{etc}. We are, however, aiming to  generalize the Einstein $A$ and $B$ coefficients in connection with the quantum Rabi oscillation under single-photon transitions. This allows us to study the novel features of the population dynamics by generalizing Einstein's rate equation with time-dependent coefficients for the two-level system. The novel features would be significant for studying non-perturbative quantum nonequilibrium statistical mechanics for the time-dependent Markovian process undergone on a cold gas of atoms or molecules.

The 3-D multimode J-C model result for the probability of stimulated or spontaneous emission a photon of (angular) frequency $\omega_{\vec{k}}$, wavevector $\vec{k}$ and polarization $s$ over $n$ such photons at time $t=0$ from a two-level system having the Bohr frequency $\omega_0=(E_2-E_1)/\hbar$ found initially ($t=0$) in the excited state in a cavity, takes the form within the dipole approximation\footnote{If the dimensions of the two-level system are small in comparison with the wavelength of the field and the wave functions of different two-level systems do not overlap, then only we can apply the dipole approximation ($\text{e}^{i\vec{k}\cdot\vec{r}_0}\simeq1$). Position ($\vec{r}_0$) of the two-level system in the cavity is not important within the dipole approximation}, as \cite{Jaynes,Agarwal,Lahiri}
\begin{eqnarray}\label{eqn:1}
P_{2\rightarrow1}^{n\rightarrow n+1}(g_{\vec{k}s}, \omega_{\vec{k}}, t)&=&4g_{\vec{k}s}^2\times(n+1)\nonumber\\&&\times\frac{\sin^2\big(\frac{\sqrt{(\omega_{\vec{k}}-\omega_0)^2+4g_{\vec{k}s}^2(n+1)} t}{2}\big)}{(\omega_{\vec{k}}-\omega_0)^2+4g_{\vec{k}s}^2(n+1)}~~~
\end{eqnarray}
where $g_{\vec{k}s}=\sqrt{\frac{\omega_{\vec{k}}}{2\hbar\epsilon_0V}}\langle\psi_1|\hat{\vec{d}}\cdot\hat{\text{e}}_{\vec{k}s}|\psi_2\rangle$ \cite{Seke}, $\hat{\vec{d}}$ is the electric dipole moment operator for the two-level system, $\hat{\text{e}}_{\vec{k}s}$ is the unit-vector for the polarization of the cavity field\footnote{Here $\hat{\text{e}}_{\vec{k}s}$ is perpendicular to $\vec{k}$.}, $n$ is the number of photons of energy $\hbar\omega_{\vec{k}}$ and mode $\vec{k}s$ each present at around the two-level system before it undergoes a spontaneous or stimulated emission resulting in $n+1$ photons of energy $\hbar\omega_{\vec{k}}$ and momentum $\hbar\vec{k}$ each after the emission and $V$ is the volume of space occupied by both the two-level system and the photons. 

Let us first consider the case of two-level system in a 3-D blackbody cavity \cite{Lahiri}. There can be infinitely large number of choices of the modes ($\vec{k}s$) of a photon for a fixed $\omega_{\vec{k}}=\omega$. This causes appearance of the density of states ($\frac{\omega^2V}{\pi^2c^3}$\footnote{It follows from $2\frac{V4\pi k^2}{(2\pi)^3}\text{d}k=2\frac{V4\pi \omega^2}{(2\pi)^3c^3}\text{d}\omega$ (for $2$ independent polarizations) where $c$ is the speed of light in the free space inside the blackbody cavity.} for two independent polarizations) once we go to description of the (angular) frequency. Thus averaging over all directions and polarizations for fixed $|\vec{k}|=\omega/c$ we get the net transition (spontaneous emission or stimulated emission) probability
\begin{eqnarray}\label{eqn:1a}
P_{2\rightarrow1}(t)&=&\sum_{n=0}^\infty\int_0^\infty p_n(\omega)P_{2\rightarrow1}^{n\rightarrow n+1}(g_{\omega}, \omega, t)\frac{\omega^2V}{\pi^2c^3}\text{d}\omega\nonumber\\&=&\sum_{n=0}^\infty\int_0^\infty p_n(\omega)P_{2\rightarrow1}^{n\rightarrow n+1}(g_{\omega}, \omega, t)\frac{\tilde{u}(\omega)d_{21}^2}{3\epsilon_02\hbar^2g_\omega^2}\text{d}\omega~~~~
\end{eqnarray}
where $g_{\vec{k}s}$ is replaced by the new coupling constant $g_{\omega}$ (such that $g_{\omega}^2=\langle g_{\vec{k}s}^2\rangle_{\text{all~directions~and~polarizations}}=\frac{\omega}{2\hbar\epsilon_0V}\frac{d_{12}^2}{3}$) once the averaging over all the directions is taken, $d_{12}=\langle\psi_1|\hat{\vec{d}}|\psi_2\rangle$ is the transition dipole moment, the factor $1/3$ comes
from averaging over all the directions of incidence and the two independent polarization states of the blackbody radiation field,  $\tilde{u}(\omega)=\frac{\hbar\omega^3}{\pi^2c^3}$\footnote{$\tilde{u}(\omega_0)$ often appears in the Planck's distribution formula and is commonly known as the ratio of the Einstein $A$ coefficient and the Einstein $B$ coefficient \cite{Hilborn}.} represents the average energy density per thermal photon per unit (angular) frequency interval, and $p_n(\omega)=(1-\text{e}^{-\frac{\hbar\omega}{k_BT}})\text{e}^{-n\hbar\omega/k_BT}$ \cite{Meekhof} is the occupation probability for $n$ thermal photons which take part in spontaneous ($n=0$) or stimulated ($n\ge1$) emission. While the transition probability $P_{2\rightarrow1}^{n\rightarrow n+1}(\omega, t)$ is sharply peaked at the resonance, the functions $\tilde{u}(\omega)/g^2_\omega$ and $p_n(\omega)$ are smooth in comparison to $P_{2\rightarrow1}^{n\rightarrow n+1}(\omega, t)$ at around the resonance. 

The transition probability $P_{2\rightarrow1}^{n\rightarrow n+1}(g_{\omega}, \omega, t)$ in Eqn.~(\ref{eqn:1a}) takes the form $P_{2\rightarrow1}^{n\rightarrow n+1}(g_{\omega}, \omega, t)\rightarrow4g_\omega^2(n+1)\pi\frac{t}{2}\delta(\omega-\omega_0)$ in the limiting case of the weak coupling constant and long time exposition ($g_\omega\ll1/t\ll\omega_0$). This result is compatible with Fermi's golden rule. The net transition probability as in Eqn.~(\ref{eqn:1a}) thus takes the form in this limiting case as $P_{2\rightarrow1}(t)\rightarrow\sum_{n=0}^\infty tp_n(n+1)\frac{\pi\tilde{u}(\omega_0)}{3\epsilon_0}d_{21}^2/\hbar^2$. Here-from one gets the rate of the emission as $|\frac{d}{dt}P_{2\rightarrow1}(t)|=R_{2\rightarrow1}(t)\rightarrow A(0)+u(\omega_0)B_{21}(0)$ where $A(0)=\frac{d_{21}^2\omega_0^3}{3\pi c^3\epsilon_0\hbar}$ is Einstein's $A$ coefficient, $B_{21}(0)=\frac{\pi d_{21}^2}{3\epsilon_0\hbar^2}$ is Einstein's $B$ coefficient and $u(\omega_0)=\tilde{u}(\omega_0)\sum_{n=0}^\infty np_n=\frac{\hbar\omega_0^3}{\pi^2c^3}\frac{1}{\text{e}^{\hbar\omega_0/k_BT}-1}$ is the average energy density of the thermal photons per unit (angular) frequency interval\footnote{The expression for $u(\omega_0)$ is often called as Planck's distribution formula.}. This is a common way of deriving Einstein coefficients from the J-C model in the weak coupling limit and long time limit \cite{Lahiri}. However, if we don't take these limits, both the time and the coupling constant would enter into the expression of the rates of the spontaneous emission and stimulated emission. Thus one can generalize the Einstein coefficients with time and coupling constant dependences. This article is dedicated to explore the time and the coupling constant dependences in the generalized Einstein coefficients and its consequences.    

The rest of this article deals with the Eqn.~(\ref{eqn:1a}). We calculate the net transition (spontaneous or stimulated emission) probability by integrating the right hand side of the Eqn.~(\ref{eqn:1a}) over the (angular) frequency $\omega$ with proper normalization for both the range ($0<\omega<\infty$) of the frequency and the distribution of the thermal photons at a temperature $T$. Then we renormalize the coupling constant ($g_{\omega}$) of the J-C model taking the quantum Rabi oscillation into account, and subsequently we generalize the Einstein coefficients towards time-dependence. Using the renormalized coupling constant we calculate the net transition probability for a lossy resonant cavity, and subsequently we discuss on the `vacuum' Rabi oscillation. Then we do the similar study of the quantum Rabi oscillations for the injected coherent field. We compare our results with the quantum Rabi oscillation data obtained by Brune \textit{at al} \cite{Brune} for various situations. Then we study the population dynamics by generalizing Einstein's rate equation with the time-dependent rate coefficients. We get entropy production of the two-level system from the population dynamics. Finally, we discuss and conclude.

\section{Jaynes-Cummings model result for the net transition probability}
Since most of the contributions in the net transition probability in Eqn. (\ref{eqn:1a}) is coming from around the resonance ($\omega\rightarrow\omega_0$), we can safely replace $\tilde{u}(\omega)$ by $\tilde{u}(\omega_0)$, $p_n(\omega)$ by $p_n(\omega_0)$ and $g_\omega$ by $g_{\omega_0}$ while integrating over $\omega$ in the domain $0<\omega<\infty$ or alternatively integrating over the generalized $n$-photon Rabi frequency $\Omega_n=\pm\sqrt{(\omega-\omega_0)^2+4g_{\omega_0}^2(n+1)}$ from -$\sqrt{\omega_0^2+4g_{\omega_0}^2(n+1)}$ to -$2g_{\omega_0}\sqrt{n+1}$ and $2g_{\omega_0}\sqrt{n+1}$ to $\infty$ as $\omega$ varies from $0$ to $\omega_0$ and $\omega_0$ to $\infty$ respectively with an avoided crossing at $\omega=\omega_0$. The first part of the integrations takes a closed form and becomes equal to the second part if we send the lower limit -$\sqrt{\omega_0^2+4g_{\omega_0}^2(n+1)}$ to $-\infty$ within the rotating wave approximation ($\omega_0^2\gg4g^2_{\omega_0}$\footnote{The rotating wave approximation $\omega+\omega_0\gg|\omega-\omega_0|$ ($\forall~\omega$) implies $\omega_0^2\gg4g^2_{\omega_0}$ to hold near the resonance.}). Thus we recast Eqn. (\ref{eqn:1a}), as
\begin{eqnarray}\label{eqn:2}
P_{2\rightarrow1}(t)&\simeq&A(0)\sum_{n=0}^\infty p_n(\omega_0)(n+1)\nonumber\\&&\times~_1F_2\bigg(\{\frac{1}{2}\}, \{1,\frac{3}{2}\}, -\frac{(\omega_nt)^2}{4}\bigg)t
\end{eqnarray}
where $~_1F_2$ is a generalized hypergeometric function expressed in the usual notation~\footnote{$_1F_2\big(\{1/2\}, \{1, 3/2\}, -(1/4) a^2 x^2\big)=\frac{1}{x}\int J_0(a x)\text{d}x$} and $\omega_{n}=2g_{\omega_0}\sqrt{n+1}$ is the $n$-photon Rabi frequency. The generalized hyper geometric function reaches $1$ exhibiting the expected result $P_{2\rightarrow1}(t)\rightarrow[A(0)+u(\omega_0)B_{21}(0)]t$  in the weak coupling limit ($g_{\omega_0}t\ll1$) and long time limit ($\omega_0t\gg1$) \cite{Lahiri}. 

\subsection{Renormalization of the coupling constant for thermal photons in a blackbody cavity}
The requirement that, $P_{2\rightarrow1}(t)$ in Eqn.~(\ref{eqn:2}) reaches $1/2$ as $t$ goes to infinity (which has also been experimentally observed \cite{Brune}), renormalizes $g_{\omega_0}$ to be the effective (or renormalized) coupling constant, as 
\begin{eqnarray}\label{eqn:2a0}
g_{\omega_0}'(\bar{n})=\frac{A(0)}{\bar{n}}\text{Li}_{-\frac{1}{2}}\big(\frac{\bar{n}}{1+\bar{n}}\big)
\end{eqnarray}
where $\bar{n}=\sum_0^\infty n~p_n(\omega_0)=\frac{1}{\text{e}^{\hbar\omega_0/k_BT}-1}$ is the average number of thermal photons in the blackbody cavity at the temperature $T$ and $\text{Li}_j(x)=x+x^2/2^j+x^3/3^j+...$ is the poly-Logarithmic function of order $j$. The real function $\text{Li}_j(x)$ though is defined for $x<1$ $\forall~j$, its special form $\text{Li}_{-\frac{1}{2}}(\frac{\bar{n}}{1+\bar{n}})$ is defined for all finite values of $\bar{n}$. The renormalized coupling constant $g_{\omega_0}'(\bar{n})$, which takes the light-matter coupling for both the thermal photons and no photons (i.e. vacuum) into account, reaches the Einstein $A$ coefficient $A(0)$ at $T\rightarrow0$. We show the same in the inset of the figure \ref{fig1}. The net transition probability in Eqn.~(\ref{eqn:2}) is a quasi-periodic function of time and has the quasi (angular) frequency $\omega_\gamma=2g_{\omega_0}\sqrt{\bar{n}+1}$ which can also be renormalized with the effective coupling constant $g_{\omega_0}'(\bar{n})$, as $\Omega_R(\bar{n})=2g_{\omega_0}'(\bar{n})\sqrt{\bar{n}+1}$. This renormalized frequency is the Rabi flopping frequency of the two-level system in the thermal radiation field\footnote{Connection of this form of the Rabi frequency with the low $\bar{n}$ will be shown below Eqn.~(\ref{eqn:2b}).}. One can, however, determine the value of the $A$ coefficient using the relation $\Omega_R(\bar{n})=2g_{\omega_0}'(\bar{n})\sqrt{\bar{n}+1}$ from the experimental data of $\Omega_R=2\pi\times47\times10^3~\text{Hz}\simeq0.295310\times10^6~$Hz \cite{Brune}.

\subsection{Generalization of the Einstein coefficients towards time-dependence under the quantum Rabi oscillation}
It is to be mentioned that, the rate of transitions ($R_{2\rightarrow1}(t)=|\frac{\text{d}}{\text{d}t}P_{2\rightarrow1}(t)|$) of the two-level system at $t\rightarrow0$ can be directly obtained from Eqn.~(\ref{eqn:2}) without referring to the Fermi's golden rule as $R_{2\rightarrow1}(0)=A(0)\sum_{n=0}^\infty p_n(\omega_0)(n+1)=B_{21}(0)~u(\omega_0)+A(0)$. If the time-derivative ($\frac{\text{d}}{\text{d}t}P_{2\rightarrow1}(t)$) be negative, then it represents the rate of transitions in the reverse order. Thus we have defined the rate of transitions with the absolute value. While the rate $R_{2\rightarrow1}(0)$ reaches $u(\omega_0)$ times the Einstein $B$ coefficient in absence of the vacuum fluctuations, it reaches the Einstein $A$ coefficient in absence of the thermal photons. The coupling constant $g_{\omega_0}$ in Eqn.~(\ref{eqn:2}) further has to be replaced by the renormalized coupling constant $g_{\omega_0}'(\bar{n})$ to ensure $P_{2\rightarrow1}(\infty)=1/2$. Eqn.~(\ref{eqn:2}) with $g_{\omega_0}$ replaced by $g_{\omega_0}'(\bar{n})$ thus unifies both the Dirac's theory of stimulated emission and the Weisskopf-Wigner theory of spontaneous emission in a single framework of the J-C model for multimodes. Such a unification was previously done only for the resonant frequency ($\omega\rightarrow\omega_0$) by the use of the Fermi's golden rule on the time-derivative of the transition probability $P_{2\rightarrow1}^{n\rightarrow n+1}(\omega, t)$ \cite{Lahiri}. Our consideration of the frequency broadening ($\triangle\omega\sim\Omega_R/2$) of the transition probability $P_{2\rightarrow1}^{n\rightarrow n+1}(\omega, t)$ around the resonant frequency ($\omega=\omega_0$) generalizes the previous unification by bringing time-dependence in the rate of the transitions $R_{2\rightarrow1}(t)=|\frac{\text{d}}{\text{d}t}P_{2\rightarrow1}(t)|$ as one can expect the same from the experimental observation of $P_{2\rightarrow1}(t)$ \cite{Brune}.

It is clear from the Eqn.~(\ref{eqn:2}) that, the stimulated emission part ($n$ of $(n+1)$) and the spontaneous emission part ($1$ of $(n+1)$) though are primarily independent in the short time scale, are secondarily dependent on each other through the $_1F_2$ part of the net transition probability as time goes on. This is possible because spontaneously emitted photon can also take part in the stimulated emission. Eventually both the spontaneous emission part and the stimulated emission part of the transition rate $R_{2\rightarrow1}(t)=|\frac{\text{d}}{\text{d}t}P_{2\rightarrow1}(t)|$ become secondarily hybrid. Thus we get the emission rate $R_{2\rightarrow1}(t)=|\frac{\text{d}}{\text{d}t}P_{2\rightarrow1}(t)|$ for the renormalized coupling constant $g_{\omega_0}'(\bar{n})$ as $R_{2\rightarrow1}(t)=\tilde{u}(\omega_0)\bar{n}B_{21}(t)+A(t)=B_{21}(t)u(\omega_0)+A(t)$ such that
\begin{eqnarray}\label{eqn:2a}
B_{21}(t)&=&B_{21}(0)\sum_{n=0}^\infty n~p_n(\omega_0)|J_0(2g_{\omega_0}'(\bar{n})\sqrt{n+1}t)|/\bar{n}\nonumber\\&\simeq& B_{21}(0)|J_0(\Omega_R(\bar{n})t)|
\end{eqnarray}
and
\begin{eqnarray}\label{eqn:2b}
A(t)&=&A(0)\sum_{n=0}^\infty p_n(\omega_0)|J_0(2g_{\omega_0}'(\bar{n})\sqrt{n+1}t)|\nonumber\\&\simeq& A(0)|J_0(\Omega_R(\bar{n})t)|
\end{eqnarray}
where $\Omega_R(\bar{n})\simeq 2g_{\omega_0}'(\bar{n})\sqrt{\bar{n}+1}$ is the Rabi flopping frequency as defined before for low photon number fluctuation ($\triangle n=\sqrt{\bar{n}}\sqrt{\bar{n}+1}\lnsim1$ \cite{Pathria}) for low $\bar{n}$. The transition rate $R_{2\rightarrow1}(t)$ becomes time-dependent along with its stimulated emission part $B_{21}(t)u(\omega_0)$ and the spontaneous emission part $A(t)$ only for the  nonzero values of the Rabi frequency $\Omega_R$. We can call $B_{21}(t)$ as the generalized Einstein $B_{21}$ coefficient and $A(t)$ as the generalized Einstein $A$ coefficient. We can also have $B_{12}(t)$ as the generalized Einstein $B_{12}$ coefficient. The generalized Einstein coefficients, however, become the original time-independent Einstein coefficients for $\Omega_R\rightarrow0$ i.e. for the case of no Rabi flopping \cite{Einstein,Dirac,Weisskopf}. It should also be mentioned that, the generalized Einstein $B_{21}$ coefficient as shown in Eqn.~(\ref{eqn:2a}) takes the form similar to that obtained in the semiclassical Rabi model \cite{Islam}. The semiclassical Rabi model, however, can not generalize the Einstein $A$ coefficient.  

On the other hand, if $P_{1\rightarrow2}^{n+1\rightarrow n}(g_\omega, \omega, t)$ be the transition probability counter to $P_{2\rightarrow1}^{n\rightarrow n+1}(g_\omega, \omega, t)$ for the (stimulated) absorption of $1$ photon from $n+1$ photons of frequency $\omega$ each and $P_{1\rightarrow2}(t)$ be the corresponding net transition probability counter to $P_{2\rightarrow1}(t)$, then we must have $P_{2\rightarrow1}^{n\rightarrow n+1}(\omega, t)+P_{1\rightarrow2}^{n+1\rightarrow n}(\omega, t)=1$ and $P_{2\rightarrow1}(t)+P_{1\rightarrow2}(t)=1$. Here-from we can show that, the rate of the transition probability for the absorption of the two-level system in presence of the average thermal photons $\bar{n}$ and $1$ emitted photon at any arbitrary time is $R_{1\rightarrow2}(t)=|\frac{\text{d}}{\text{d}t}P_{1\rightarrow2}(t)|=\tilde{u}(\omega_0)(\bar{n}+1)B_{12}(t)$. This relation leads to the equality $B_{12}(t)=B_{21}(t)$ as because $\tilde{u}(\omega_0)B_{21}(t)=A(t)$ holds for any arbitrary time according to Eqns.~(\ref{eqn:2a}) and (\ref{eqn:2b}). 

\begin{figure}
\includegraphics[width=0.98 \linewidth]{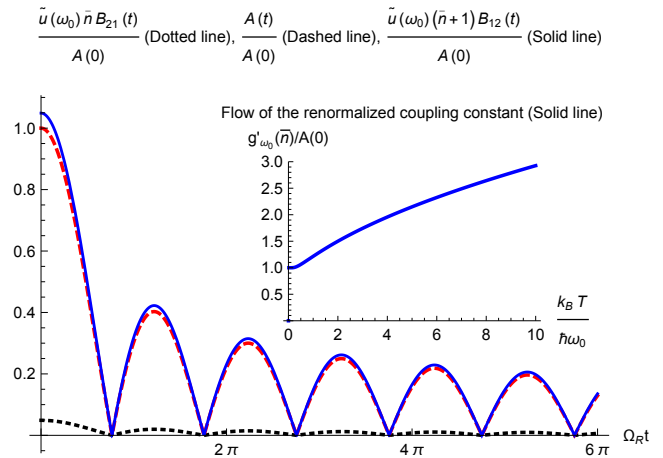}
\caption{Multimode J-C model results for generalized Einstein coefficients for $\omega_0=2\pi\times51.099\times10^9$ Hz and $T=0.8$ K as taken in Ref.\cite{Brune} for the circular Rydberg states (with the principal quantum number $n=50$ and $n=51$) of a $^{87}$Rb atom. While the solid and the dotted lines follow Eqn.~(\ref{eqn:2a}), the dashed line follows Eqn.~(\ref{eqn:2b}). Envelope of the dashed lines follow $A(t)/A(0)\equiv\sqrt{2/\pi\Omega_R t}$ for large $t$. Solid line of the inset represents temperature dependence of the renormalized coupling constant and follows Eqn. (\ref{eqn:2a0}).
\label{fig1}}
\end{figure}

We plot all the generalized Einstein coefficients with proper weightage for stimulated emission rate (dotted line), spontaneous emission rate (dashed line) and absorption rate (solid line) all in units of $A(0)$ in the figure \ref{fig1} for a low temperature $T=0.8$ K so that the background of two-level system is filled with a very small number of average thermal photons ($\bar{n}=0.0489$ \cite{Brune}). While on average $\bar{n}$ thermal photons are present in the background of the two-level system for its stimulated emission, on average $\bar{n}$ thermal photons and one emitted photon are present in the background of two-level system for its absorption. This makes significant difference between the two processes corresponding to the observation of the `vacuum'\footnote{We are calling it to be `vacuum' because the background of the two-level system in the cavity is truly not empty at $T=0.8$ K.} Rabi oscillation at the low temperature ($T=0.8$ K) by Brune \textit{et al} \cite{Brune}. It is clear from the figure \ref{fig1} that, the `vacuum' Rabi oscillation takes place due to subsequent interplay of the spontaneous emission and absorption. Role of the stimulated emission is suppressed in the `vacuum' Rabi oscillation. On the other hand, role of the spontaneous emission is suppressed at a higher temperature. In that case, the dotted line and the solid line in the figure \ref{fig1} would come close to each other.  

\subsection{Renormalization of the coupling constant for photons in a lossy resonant cavity}
Let us now consider the spontaneous or stimulated emission from the two-level system in a lossy resonant cavity, say a Fabry-Perot cavity, with $z$-axis be the cavity axis \cite{Brune,Raimond}. Above result for the blackbody cavity is expected to be unaltered if the separation of the two reflecting walls of the resonant cavity is several times larger than the wavelength of the resonant mode. The quantum Rabi oscillation needs the emitted photon to have higher life time ($\sim 200~\mu$s) than that in the free space, so that it can be repeatedly reflected by the cavity mirrors before it actually leaks out of the cavity resulting in loss (leakage) through the holes (of size $\sim$mm$^2$ each) on the cavity axis or becomes absorbed (or scattered) in the walls of the cavity resulting in the ‘‘Ohmic’’ loss \footnote{A.E. Siegman, \textit{Lasers}, University Science Books, Sausalito, sec. 7.2, p. 267 and sec. 8.3, p. 323 (1986)}. The parameter which ensures the higher life time is the higher mode quality factor ($Q=7\times10^7$ \cite{Brune}) of the cavity. However, there is additional loss as because the curved surface of the cylindrical geometry of the cavity is open. Thus the probability that the emitted photon escapes from the cavity through the curved surface of the cylindrical shaped open cavity (of circular mirrors of radius $r$ each and separation $h$) is $p_0=\frac{2\pi rh}{2\pi rh+2\pi r^2}=\frac{1}{1+\frac{r}{h}}$ which results the net quality factor as $Q'=\frac{1}{\frac{1}{Q}+\frac{p_0A(0)}{\omega_0}}$\footnote{Here $A(0)$ is the frequency broadening ($\triangle\omega$ around the resonance frequency $\omega_0$) for the natural decay in the free space \cite{Weisskopf}. The natural decay in the free space results the $Q$-factor $\frac{\omega_0}{\triangle\omega}=\frac{\omega_0}{A(0)}$.}. Since the individual loss leads to a Lorentzian distribution, convolution of the above two losses (in the short time scale) also leads to the Lorentzian distribution $u'(\omega)=u(\omega_0)\frac{2}{\pi}\frac{(\omega_0/Q')^2}{4(\omega-\omega_0)^2+(\omega_0/Q')^2}$ with the net width $\omega_0/Q'$ over the Planck's distribution $u(\omega)\simeq u(\omega_0)$. However, above form of the net transition probability (Eqn.~(\ref{eqn:2})) would be unaltered if the broadening due to the losses is much higher than that due to the natural decay (i.e. $\omega_0/Q'\gg A(0)$). Thus, we recast Eqn.~(\ref{eqn:2}) by further renormalizing the coupling constant as shown in Eqn.~(\ref{eqn:2a0}), as
\begin{eqnarray}\label{eqn:3}
P_{2\rightarrow1}(t)&\simeq&A(0)\sum_{n=0}^\infty p_n(\omega_0)(n+1)\nonumber\\&&\times~_1F_2\bigg(\{\frac{1}{2}\}, \{1,\frac{3}{2}\}, -[g_{\omega_0}'(\bar{n})\sqrt{n+1}t]^2\bigg)t.~~~~
\end{eqnarray}
Incidentally we have $A(0)\simeq15.6765~$Hz in free space or in a (very large) blackbody cavity for $d_{21}=1250a_0e$ \cite{Raimond} of the two-level system ($^{87}$Rb) of our interest and $\omega_0/Q'\simeq250210$ Hz for the cavity of our interest \cite{Brune}. Hence the condition $\omega_0/Q'\gg A(0)$ is well met if the value of $A(0)$ remains same (or decreases) in the cavity space. Otherwise, each (angular) frequency in the net transition probability (Eqn.~(\ref{eqn:3})) would have to be weighted by the Lorentzian distribution $u'(\omega)$.  

\subsection{Quantum Rabi oscillations for the two-level system in a lossy resonant cavity}
However, value of the Einstein $A$ coefficient ($A(0)$) increases enormously in the Fabry-Perot cavity due to the Purcell effect \cite{Purcell}. Broadening due to the losses may not be so large in comparison to $A(0)$ in this situation. Each frequency in the net transition probability (Eqn.~(\ref{eqn:3})) should be weighted by the Lorentzian distribution $u'(\omega)=u(\omega_0)\frac{2}{\pi}\frac{(\omega_0/Q')^2}{4(\omega-\omega_0)^2+(\omega_0/Q')^2}$ in this case. Thus Eqn.~(\ref{eqn:3}) would be further recast in a similar way of reaching Eqn.~(\ref{eqn:2}) from Eqn.~(\ref{eqn:1}), as  
\begin{eqnarray}\label{eqn:4a}
P_{2\rightarrow1}(t)&=&A(0)\sum_{n=0}^\infty p_n(\omega_0)[n+1]\frac{4}{\pi}\times\nonumber\\&&\int_{\omega_{n}}^\infty\frac{(\omega_0/Q')^2}{4(\Omega_{n}^2-\omega_{n}^2)+(\frac{\omega_0}{Q'})^2}\frac{\sin^2(\Omega_{n}t/2)}{\Omega_{n}\sqrt{\Omega_{n}^2-\omega_{n}^2}}\text{d}\Omega_{n}.~~~~
\end{eqnarray}
where $\omega_{n}=2g_{\omega_0}'(\bar{n},Q')\sqrt{n+1}$ is the new renormalized $n$-photon Rabi frequency and $g_{\omega_0}'(\bar{n},Q')$ is the new renormalized coupling constant which is to be  determined by setting the limit $P_{2\rightarrow1}(\infty)=1/2$. 

\subsubsection{`Vacuum' Rabi oscillation}
Eqn.~(\ref{eqn:4a}) would be approximated by further neglecting the photon-number fluctuation ($\triangle n=\sqrt{\bar{n}}\sqrt{1+\bar{n}}$ \cite{Pathria}) at the higher order of the Taylor expansion of $P_{2\rightarrow1}(t)$ about $n=\bar{n}$ for low $\bar{n}$ ($\bar{n}\lnsim1$) at a low temperature\footnote{Here $\bar{n}$ is a small quantity at a low temperature. Thus $\sqrt{n+1}$ at the argument of the generalized hepergeometric function in Eqn.~(\ref{eqn:3}) is approximated as $\sqrt{\bar{n}+1}$ $\forall~n$.}, as 
\begin{eqnarray}\label{eqn:4}
P_{2\rightarrow1}(t)&\simeq&A(0) [\bar{n}+1]\frac{4}{\pi}\times\nonumber\\&&\int_{\omega_{\bar{n}}}^\infty\frac{(\omega_0/Q')^2}{4(\Omega_{\bar{n}}^2-\omega_{\bar{n}}^2)+(\frac{\omega_0}{Q'})^2}\frac{\sin^2(\Omega_{\bar{n}}t/2)}{\Omega_{\bar{n}}\sqrt{\Omega_{\bar{n}}^2-\omega_{\bar{n}}^2}}\text{d}\Omega_{\bar{n}}.~~~~
\end{eqnarray}
The number $1$ next to $\bar{n}$ in Eqn.~(\ref{eqn:4}) arises purely from the quantum fluctuations. Effect of the quantum fluctuations are suppressed in the classical regime ($k_BT/\hbar\omega_0\gg1$). Thus Eqn.~(\ref{eqn:4}) corresponds to the classical Rabi oscillation for $\bar{n}\gg1$. However, the new renormalized coupling constant used in Eqn.~(\ref{eqn:4}) can be determined by setting the limit $P_{2\rightarrow1}(\infty)=1/2$ as $g_{\omega_0}'(\bar{n},Q')=\frac{A(0)\sqrt{\bar{n}+1}}{1+4g_{\omega_0}'(\bar{n},Q')\sqrt{\bar{n}+1}Q'/\omega_0}$ which further determines the Rabi frequency for a lossy resonant cavity at a low temperature as $\Omega_R(\bar{n},Q')=2g_{\omega_0}'(\bar{n},Q')\sqrt{\bar{n}+1}=\frac{-1+\sqrt{1+16A(0)[\bar{n}+1]Q'/\omega_0}}{4Q'/\omega_0}$. This relation further determines the Einstein $A$ coefficient to be as $A(0)=\frac{\Omega_R}{2({\bar{n}+1})}+\frac{\Omega_R^2Q'}{\omega_0({\bar{n}+1})}$. While the 1st term of $A(0)$ represents the Einstein $A$ coefficient in the free space, the 2nd term represents enhancement of the $A$ coefficient due to the Purcell effect in the resonant cavity. Now we get enhanced value of the $A$ coefficient as $A(0)\simeq0.473053\times10^6$Hz for the $^{87}$Rb atom in the resonant cavity of our interest \cite{Brune}. The net quality factor corresponding to this $A(0)$ now takes the value $Q'=1.28318\times10^6$. 

\begin{figure}
\includegraphics[width=0.98 \linewidth]{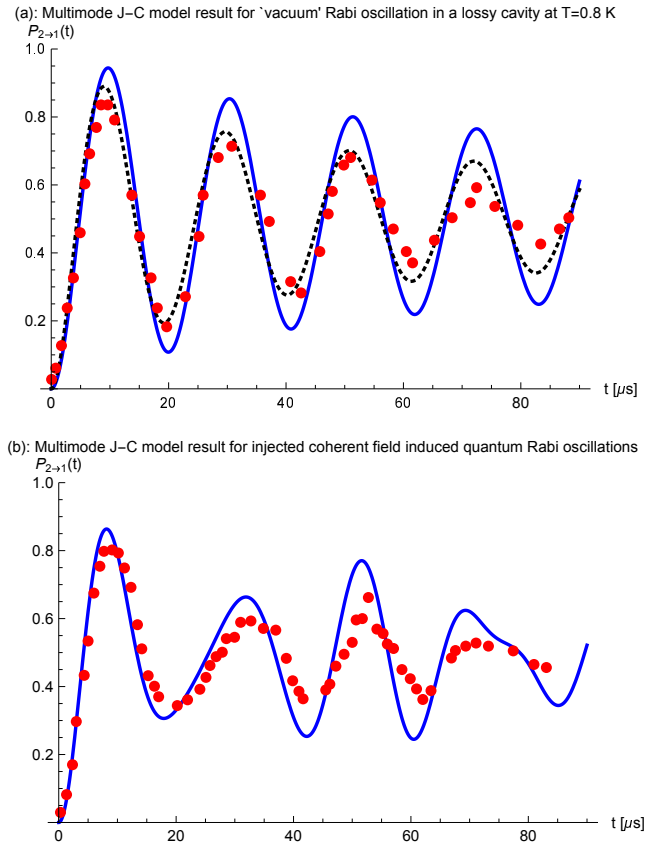}
\caption{
(a): Solid line represents the `vacuum' Rabi oscillation, and follows Eqn.~(\ref{eqn:4}) for the Bohr frequency $\omega_0=2\pi\times51.099\times10^9~$Hz, Rabi frequency $\Omega_R=2\pi\times47\times10^3~$Hz and average number of thermal photons $\bar{n}=0.0489$. Circles represent corresponding experimental data \cite{Brune} adapted for the circular Rydberg states (with the principal quantum number $n=50$ and $n=51$) of $^{87}$Rb atoms in an open resonant cavity of the Q-factor $Q=7\times10^7$ and size $\pi(50/2)^2\times27$ mm$^3$ at the temperature $T=0.8$ K. The dotted line represents the same for $Q=7\times10^5$.\\
(b): Solid line represents injected coherent field induced quantum Rabi oscillation, and follows Eqn.~(\ref{eqn:4a}) for the same parameters as mentioned above except for Rabi frequency $\Omega_R=2\pi\times55.6949\times10^3~$Hz~ and the average number of injected photons $\bar{n}=0.4$ in the lossy cavity. Circles represent corresponding experimental data \cite{Brune} adapted for the same two-level system.
\label{fig1b}}
\end{figure}

We plot the right hand side of the Eqn.~(\ref{eqn:4}) in the figure \ref{fig1b}-a for the $^{87}$Rb atom in the resonant cavity \cite{Brune}. The solid line in the figure \ref{fig1b}-a represents the `vacuum' Rabi oscillation in the resonant cavity for the parameters as mentioned in the figure-caption. The cavity is truly not empty rather has on the average $\bar{n}=0.0489$ thermal photons in it \cite{Brune}. The dotted line represents a fit with the same equation but for a lower value ($Q=7\times10^5$) of the $Q$-factor. Damping of the Rabi oscillation even in the cavity, as shown in figure \ref{fig1b}-a, is caused due to the finite width ($\sim\Omega_R/2$) of the frequency distribution around the resonance. Better matching for the lower $Q$-factor can be attributed to the substantial losses from the resonant cavity corresponding to the frequency broadening due to inhomogeneous light-matter coupling along the cavity axis \cite{Brune}, Doppler broadening due to the speed distribution of $^{87}$Rb atoms in the cavity, thermal broadening, \textit{etc}. 

\subsubsection{Quantum Rabi oscillations for injected coherent field}
Quantum Rabi oscillation was not only observed in the form of `vacuum' Rabi oscillation but also in the form of coherent field induced Rabi oscillations \cite{Brune}. Let us also do similar study of injected coherent field induced quantum Rabi oscillations. The main difference in this respect comes from the probability distribution of photons. While the probability distribution follows exponential law for the thermal photons, it follows Poisson distribution $p_{n}(\omega_0)=\frac{\bar{n}^n}{n!}\text{e}^{-\bar{n}}$ for $\bar{n}$ average number photons of frequency $\omega_0$ in the coherent field. Averaging of the transition probability $P_{2\rightarrow1}(t)$ over the polarization is no longer needed. Thus the factor $\frac{1}{3}$ is no longer needed in Eqn.~(\ref{eqn:1a}). All values of the frequencies as mentioned in Eqn. (\ref{eqn:1a}) are also not welcome in the injected coherent field as because the coherent field has finite maximum detuning, say $(\omega-\omega_0)_{max}=10^6$~Hz in Brune \textit{et al}s' experiment \cite{Brune}. However, significant contribution in the integrals in Eqns.~(\ref{eqn:1a}) and (\ref{eqn:4a}) are coming from the domain $\omega_0-\Omega_R$ to $\omega_0+\Omega_R$ of the frequency $\omega$. Incidentally, the maximum detuning is about $3$ times of $\Omega_R$ \cite{Brune}. Thus Eqn.~(\ref{eqn:4a}) with $p_{n}(\omega_0)=\frac{\bar{n}^n}{n!}\text{e}^{-\bar{n}}$ would still be okay for the net transition probability of the spontaneous emission or the stimulated emission induced by the injected coherent field. Injected coherent field increases the light-matter coupling constant so as the Rabi frequency. The renormalized coupling constant ($g_{\omega_0}'(\bar{n},Q')$) can be determined from the limiting value $P_{2\rightarrow1}(\infty)=1/2$ for the previous values of $A(0)$ ($=~0.473053\times10^6$ Hz) and $Q'$ ($=~1.28318\times10^6$ Hz)). We determine the new renormalized coupling constant to the second order in experimental value $\bar{n}=0.4$ ($\lnsim1$) \cite{Brune} as $g_{\omega_0}(0.40,Q')\simeq0.147877\times10^6$ Hz. Here-from we get the Rabi frequency as $\Omega_R=2g_{\omega_0}(0.40,Q')\sqrt{0.4+1}\simeq55.6949\times2\pi\times10^3$~Hz. 

The solid line in the figure \ref{fig1b}-b represents the injected coherent field induced quantum Rabi oscillations in the resonant cavity for the parameters as mentioned in the figure-caption. Amplitude of the quantum Rabi oscillations in the figure \ref{fig1b}-b is observed to be less than that in the figure \ref{fig1b}-a because of the larger photon number fluctuation in the case of the figure \ref{fig1b}-b. Photon number fluctuation kills the quantum Rabi oscillations for large values of $\bar{n}$ ($\gg1$). Damping of the Rabi oscillations even for the coherent field in the cavity, as shown in figure \ref{fig1b}-b, is caused due to the finite width of the frequency broadening at around the resonance. However, we see good agreement of our theoretical result with the experimental data in the figure \ref{fig1b}-b. Matching would have been better had we considered additional substantial losses due to the inhomogeneous light–matter coupling, Doppler broadening, thermal broadening, higher order effect of $\bar{n}$ in the light-matter coupling constant, \textit{etc}. We could show the collapse and the revival \cite{Eberly} well for large values of $\bar{n}\gtrsim0.85$ \cite{Brune} had we known the net quality factor ($Q'$) of the resonant cavity. The net quality factor would significantly decrease for this case of the injected coherent field due to more losses from the cavity.

\section{Jaynes-Cummings model result for population dynamics with Einstein's rate equations}
While the generalized Einstein $B$ coefficients are same and time-dependent in the semiclassical Rabi model, the Einstein $A$ coefficient remains the original time-independent $A$ coefficient in the same model \cite{Islam}. This causes even a small $\Omega_R$ to greatly influence the time-evolution of the statistical mechanical occupation probabilities $P_1(t)$ and $P_2(t)$ of the states $|\psi_1\rangle$ and $|\psi_2\rangle$, respectively \cite{Islam}. However, we already have all the generalized Einstein coefficients to be time-dependent in a similar fashion. Let us now investigate how the occupation probabilities evolve with time for the multimode J-C model results of the generalized Einstein coefficients as obtained in Eqns.~(\ref{eqn:2a}) and (\ref{eqn:2b}). 

\begin{figure}
\includegraphics[width=0.98 \linewidth]{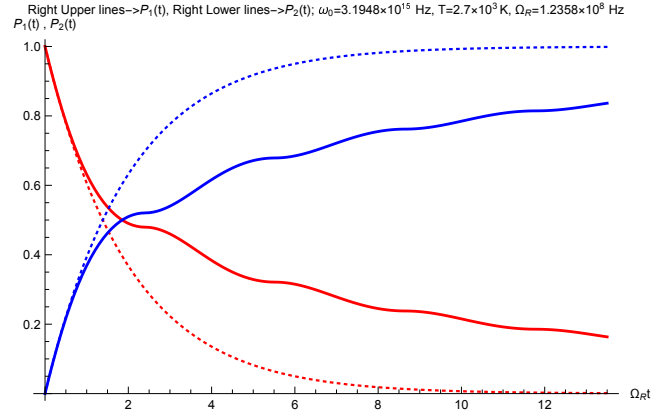}
\caption{
Occupation probabilities for the $3s_{\frac{1}{2}}$ and $3p_{\frac{1}{2}}$ states of an $^{23}$Na atom in the thermal radiation field with the condition that the system initially was in the upper level. Right lower and right upper solid lines follow Eqns.~(\ref{eqn:13}) and (\ref{eqn:14}), respectively for the parameters as mentioned in the figure corresponding to $d_{21}=2.5ea_0=2.1196\times10^{-29}$Cm~\cite{Boyd}. Lower and upper dotted lines represent Einstein probabilities for the same system, and follow Eqn.~(\ref{eqn:12b}) and its complementary, respectively. 
\label{fig2}}
\end{figure}

Time-evolution of the occupation probabilities are to be determined from Einstein's rate (master) equations~\cite{Einstein,Griffiths,Feynman} which are now revised with the generalized Einstein coefficients in Eqns.~(\ref{eqn:2a}) and (\ref{eqn:2b}), as
\begin{eqnarray}\label{eqn:10}
\frac{d P_2}{dt}&=&-A(t) P_2(t)-u(\omega_0)B_{21}(t)P_2(t)\nonumber\\&&+u(\omega_0)B_{12}(t)P_1(t)
\end{eqnarray}
and
\begin{eqnarray}\label{eqn:11}
\frac{d P_1}{dt}&=&A(t) P_2(t)+u(\omega_0)B_{21}(t)P_2(t)\nonumber\\&&-u(\omega_0)B_{12}(t)P_1(t)
\end{eqnarray}
where we also have $B_{12}(t)=B_{21}(t)$  as discussed below Eqn.~(\ref{eqn:2b}). Time-evolution of the occupation probabilities, because of the constraint $P_1(t)+P_2(t)=1$, can be solely determined from any one of the above two equations, say Eqn.~(\ref{eqn:10}), with $P_1(t)$ be replaced by $1-P_2(t)$. Thus, we recast Eqn.~(\ref{eqn:10}) with the spontaneous emission rate $A(t)$ and the stimulated emission rate $R(t)=B_{21}(t)u(\omega_0)=B_{12}(t)u(\omega_0)$, as
\begin{eqnarray}\label{eqn:12a}
\frac{d P_2}{dt}=R(t)-[A(t)+2R(t)]P_2(t).
\end{eqnarray}
While $R(0)$ is the rate-coefficient for stimulated emission/absorption found within the first order time-dependent perturbation theory of quantum mechanics \cite{Dirac}, $A(0)$ is the rate-coefficient for the spontaneous emission found within the first order perturbation theory\footnote{The first order perturbation theory is compatible with Fermi's golden rule.} of quantum electrodynamics \cite{Weisskopf}. Eqn.~(\ref{eqn:12a}) has a physical solution for $R(t)=R(0)$ and $A(t)=A(0)$ with the initial condition $P_2(0)=1$, as~\cite{Einstein,Griffiths}
\begin{eqnarray}\label{eqn:12b}
P_2(t)&=&\frac{R(0)}{A(0)+2R(0)}\nonumber\\&&+\bigg[1-\frac{R(0)}{A(0)+2R(0)}\bigg]\text{e}^{-[A(0)+2R(0)]t}
\end{eqnarray}
which is often equated with the (time-independent) Boltzmann probability $P_2(\infty)=\frac{\text{e}^{-E_2/k_BT}}{\text{e}^{-E_1/k_BT}+\text{e}^{-E_2/k_BT}}$ in thermal equilibrium for $t\rightarrow\infty$~\cite{Einstein,Griffiths}. Occupation probability of the lower level, on the other hand, can be given by $P_1(t)=1-P_2(t)$. Eqn.~(\ref{eqn:12b}) is Einstein's semiclassical result for the occupation probability \cite{Einstein,Griffiths}. Let us call the time-dependent probabilities $P_1(t)$ and $P_2(t)$ which follow from Eqn.~(\ref{eqn:12b}), as Einstein probabilities \cite{Einstein,Griffiths}. Dotted lines in figure \ref{fig2} represent the Einstein probabilities. It is clear from Eqns.~(\ref{eqn:2a}) and (\ref{eqn:2b}) that, $R(t)=R(0)$ and $A(t)=A(0)$ are possible only when $\Omega_R\rightarrow0$ i.e. when there is no Rabi oscillation. Our aim for the rest of the article is to modify the Einstein probabilities due to the presence of the Rabi flopping in the same system within the quantum field theoretic description of the multimode J-C model. 

We solve Eqn.~(\ref{eqn:12a}) with the initial condition $P_2(0)=1$ for $B_{21}(t)=B_{12}(t)=R(t)/u(\omega_0)$ of Eqn.~(\ref{eqn:2a}) and $A(t)$ of  Eqn.~(\ref{eqn:2b}), as
\begin{eqnarray}\label{eqn:13}
P_2(t)&=&\text{e}^{-[A(0)+2R(0)]f_{\Omega_R}(t)}\bigg[1+R(0)\times\nonumber\\&&\int_0^t\text{e}^{[A(0)+2R(0)]f_{\Omega_R}(\tau)}|J_0(\Omega_R\tau)|\text{d}\tau\bigg]
\end{eqnarray}
where $f_{\Omega_R}(t)$ is given by
\begin{eqnarray}\label{eqn:13x}
f_{\Omega_R}(t)&=& _1F_2\bigg(\{\frac{1}{2}\}, \{1,\frac{3}{2}\}, -\frac{\Omega_R^2 t^2}{4}\bigg)\big[2\text{U}(J_0(\Omega_R t))-1\big]t\nonumber\\&&-\frac{2}{\Omega_R}\sum_{j=1}^{\floor*{\Omega_R t}}\bigg[(-1)^j{\gamma_{0,j}} _1F_2\bigg(\{\frac{1}{2}\}, \{1,\frac{3}{2}\}, -\frac{\gamma_{0,j}^2}{4}\bigg)\nonumber\\&&\times\text{U}(\Omega_R t-\gamma_{0,j})\bigg]
\end{eqnarray}
where $\gamma_{0,j}$ is the $j$th zero of the Bessel function ($J_0$) of the first kind of order $0$ and $\text{U}$ is the unit step function. Now, we get the occupation probability of the lower level from Eqn.~(\ref{eqn:13}), as
\begin{eqnarray}\label{eqn:14}
P_1(t)=1-P_2(t).
\end{eqnarray}
Eqns.~(\ref{eqn:13}) and (\ref{eqn:14}) are our quantum mechanical results for the occupation probabilities of the two states of the two-level system in the thermal radiation field. We plot these probabilities in the figure \ref{fig2} for the relevant values of the parameters for the $3s_{\frac{1}{2}}$ and $3p_{\frac{1}{2}}$ states of an $^{23}$Na atom. We have profusely considered the temperature to be equal to $2700$~K which is the usual temperature of the sodium vapour lamp and the usual temperature for the excitement of the two states. Rate of the stimulated emission ($R(0)=0.000059409~\Omega_R$) is much less than that of the spontaneous emission ($A(0)=0.499941~\Omega_R$) at such a temperature. This causes significant deviation of the occupation probability from the Einstein probability. Amplitude of the partial oscillation having quasi-frequency $\Omega_R/\pi$ in the occupation probability would have increased if we had taken even a lower value of $R(0)/A(0)$ ($\ll1$) at a lower temperature. Occupation probability, in contrary to that of the semiclassical Rabi model \cite{Islam}, asymptotically ($t\rightarrow\infty$) approaches the Einstein probability so as the Boltzmann probability. It is clear from the figure \ref{fig2} that, the quantum Rabi oscillation slows down the occupation probability reaching the Boltzmann probability. The deviation of the occupation probability from the Einstein probability as well as the amplitude of the partial oscillation would decrease had the ratio $R(0)/A(0)$ been taken large ($\gtrsim 1$) at a higher temperature. Our result, of course, exactly matches with the Einstein probability if the Rabi flopping is completely turned off, i.e., if we take $\Omega_R\rightarrow0$.

\section{Non-equilibrium statistical mechanical implications}

\begin{figure}
\includegraphics[width=0.98 \linewidth]{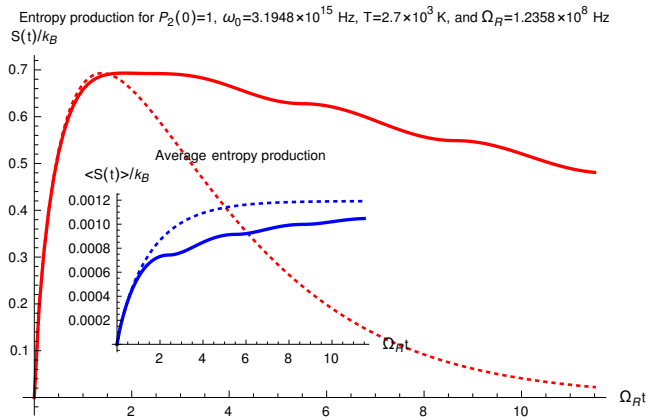}
\caption{Entropy production for the $3s_{\frac{1}{2}}$ and $3p_{\frac{1}{2}}$ states of an $^{23}$Na atom in the thermal radiation field. Plot (solid line) follows from Eqn.~(\ref{eqn:15}) for the parameters as mentioned in figure~ \ref{fig2}. Dotted line represent the same obtained from Einstein probabilities (Eqn.~(\ref{eqn:12b}) and its complementary). Solid line in the inset represents the average entropy production for same parameters except the initial condition. Dotted line in the inset represents the same based on the Einstein probabilities.
\label{fig3}}
\end{figure}

Though the light-matter interactions take place in short time scale ($t\sim1/\omega_0$), dipole-transitions take place in longer time scale ($t\sim1/\Omega_R$). Occupancy of the two levels of the system becomes probabilistic as because it is not known exactly when the two-level system makes a transition.  This loss of information leads to the entropy production of the two-level system, as \cite{Feynman,Islam}
\begin{eqnarray}\label{eqn:15}
S(t)=-k_B[P_1(t)\ln(P_1(t))+P_2(t)\ln(P_2(t))].
\end{eqnarray}
We show the time-dependence of the entropy production in the figure \ref{fig3} for the occupation probabilities (Eqns.~(\ref{eqn:13}) and (\ref{eqn:14})) and the Einstein probabilities (Eqn.~(\ref{eqn:12b}) and its complementary) for the fixed temperature $T=2700$ K and initial condition $P_2(0)=1$. We already have mentioned that, the quantum Rabi oscillation slows down the occupation probability reaching the Boltzmann probability. Similar feature is also apparent in the entropy production (solid line) in the figure \ref{fig3} where we also have plotted the entropy production (dotted line) which has been calculated based on the Einstein probabilities. The  entropy of the two-level system is always less than equal to $k_B\ln(2)$ as expected. The two forms of the entropy production eventually meet the equilibrium entropy at $t\rightarrow\infty$. It is clear from the figure \ref{fig3} that, entropy of the two-level system is not always an ever non-decreasing function of time at least for the initial condition $P_2(0)=1$.  This is, however, not an example of the violation of the second law of stochastic thermodynamics. Jarzynski equality rather allows non-increase of the entropy for some (not all) realizations of the initial conditions \cite{Jarzynski}. We also show average entropy production of two-level system in the inset of the figure \ref{fig3} for all the realizations of the initial conditions ($P_2(0)=1~\&~P_1(0)=1$) with their proper statistical weights (Boltzmann probabilities).  

It is clear from the inset of the figure \ref{fig3} that, the average entropy productions which are calculated based on both the occupation probabilities (solid line) and the Einstein probabilities (dotted line), however, are  always non-decreasing function of time at least for $R(0)/A(0)\ll1$ as well as for low photon number fluctuation ($\triangle n=\sqrt{\bar{n}}\sqrt{\bar{n}+1}\lnsim1$). Thus we validate the second law stochastic thermodynamics for a two-level system in the thermal radiation field. Hence we can safely say that, a two-level system or a gas of two-level systems in the thermal radiation field is a practical example of a thermodynamically isolated system for $R(0)/A(0)\ll1$.

\section{Conclusion}
We have obtained multimode Jaynes-Cummings model results for all the generalized Einstein coefficients for a two-level system in the thermal radiation field, and have shown that, all the generalized Einstein coefficients depend on time and Rabi frequency in a similar manner for low photon number fluctuation. These results are accurate for fairly large Bohr frequency ($\omega_0\gg\Omega_R$), and are significantly different from the results obtained within the first order time-dependent perturbation theory which considers $\Omega_R\rightarrow0$. Renormalization of the light-matter coupling done for both the vacuum field and thermal photons together leads to such a difference from the previous theories. We have obtained analytical results within the multimode Jaynes-Cummings model for the quantum Rabi oscillations for both the thermal photons and the photons of an injected coherent field. We also have studied the population dynamics for the two-level system by generalizing the Einstein rate equations with the generalized Einstein coefficients. Our results on the quantum Rabi oscillations match well with the experimental data \cite{Brune}. The population dynamics obtained by us differs significantly from that obtained in the semiclassical theory \cite{Einstein,Griffiths} for a low temperature.  Our generalization of the Einstein coefficients is an invitation to the experimentalists for direct measurement of the Einstein coefficients for the two-level system(s) in a blackbody cavity.

While the Einstein coefficients deal with the fundamental processes (e.g. spontaneous emission, stimulated emission and absorption), they don't directly deal with the Rabi oscillation. On the other hand, the quantum Rabi oscillation deals with the fundamental processes. Hence it is possible to derive Einstein coefficients from the analyses of the quantum Rabi oscillation, and we have done that for nonzero Rabi frequency. Thus we have generalized the Einstein coefficients towards time and Rabi frequency dependences.

Drexhage observed alterations in the rate of spontaneous emission, regarding the influence the atomic radiation, while working on the fluorescence of organic dyes deposited on dielectric films over a metallic mirror \cite{Drexhage,Haroche-1998}. Once a two-level system in a resonant cavity emits a photon it is periodically reabsorbed in the cavity exhibiting the quantum Rabi oscillation. Thus time-dependence of the Einstein coefficients is not a surprise at least for a two-level system in a resonant cavity. However, probability of the reabsorption is negligibly small for the same system in the free space\footnote{Here, free space refers to a large blockbody cavity.}. Thus the Einstein coefficients are not found to be time-dependent in the free space.  

We are not able to compare the result on the generalized Einstein $B$ coefficient with the existing experimental data as because they have not been obtained by any direct measurement; rather, experimentalists apply time-dependent perturbation theory ($A/B=\frac{\hbar\omega_0^3}{\pi^2c^3}$) for the indirect measurement of the Einstein $B$ coefficient from the experimental value of the Einstein $A$ coefficient \cite{Lawrence}. Time-dependence of the generalized Einstein $A$ coefficient could have been caught by the experimentalists had they measured it for longer time scale i.e. the time scale of the quantum Rabi oscillation. Measurement of the absorption coefficient \cite{Hilborn} for a two-level system doesn't also serve the purpose of capturing the time-dependence of the generalized Einstein $B$ coefficient, as because, averaging of the absorption of photons of all possible frequencies and of all possible direction of incidence is not considered in the measurement \cite{Kiselev}.        

While the quantum Rabi oscillation is studied for strong light-matter interactions ($g_{\omega_0}\gg\sup\{\gamma,~\kappa\}$\footnote{Here $\gamma$ is the non-resonant decay rate and $\kappa=\omega_0/Q$ is the photon decay rate of the cavity \cite{Fox}.} \cite{Fox}), the Einstein rate equations are often applied for weak light-matter interactions ($g_{\omega_0}\ll\sup\{\gamma,~\kappa\}$ \cite{Fox}). Incidentally, the multimode J-C model gives results in both the weak coupling regime and the strong coupling regime as far as the rotating wave approximation ($2g_{\omega_0}\ll\omega_0$) is applicable. Thus we have been interested in bridging the quantum Rabi oscillation and the phenomenological rate equations by the multimode J-C model. The partial oscillations, as shown in figure (\ref{fig2}), are expected to be damped for the broadband excitations in the intermediate regime \cite{Cohen-Tannudji2}. The experimental data \cite{Brune}, which we have compared with our results in the figure \ref{fig1b}, satisfy both the strong coupling condition and the rotating wave approximation. The J-C model, however, is not applicable in the ultrastrong coupling ($g_{\omega_0}\sim\omega_0$) and deep strong coupling ($g_{\omega_0}\gg\omega_0$) regimes \cite{Niemczyk,Forn-Diaz1}. The quantum Rabi model which generalizes the J-C model is applicable in these regimes \cite{Xie,Forn-Diaz}.  

Throughout the article by ``frequency" we have meant ``angular frequency".

We have dealt with a single two-level system (qubit) in the 3-D multimode J-C model. Transition probability calculated for this system would not have changed if we had taken non-interacting and distinguishable identical two-level systems.

Role of the fundamental processes in the time-evolution of entropy of a system are shown by considering the multimode J-C model as a toy model for the two-level system in the thermal radiation field. Time-dependence of the generalized Einstein coefficients opens a path to go beyond Pauli-von Neumann formalism of the non-equilibrium statistical mechanics \cite{Islam}. The population dynamics studied by us would be useful for studying non-perturbative quantum nonequilibrium statistical mechanics for the time-dependent Markovian process undergone on a cold gas of atoms or molecules.  

Although the limit $\Omega_R\rightarrow0$ gives all the Einstein coefficients back, yet the time-dependence of the generalized Einstein coefficients plays a significant role in the dynamics of probabilities of the two states of the system either at a low temperature or at a low average number of thermal photons. The two-level system approaches thermal equilibrium in the thermal radiation field as because the temporal parts of all the generalized Einstein coefficients are same as far as low photon number fluctuation ($\triangle n=\sqrt{\bar{n}}\sqrt{\bar{n}+1}\lnsim1$) is concerned. Had the generalized $A$ coefficient been a constant, the oscillations in the generalized $B$ coefficient even for very small $\Omega_R$, would have driven the system away from the thermal equilibrium at any finite temperature \cite{Islam}. The temporal parts are expected to be differed for large photon number fluctuation, and the two-level system is expected to go away from thermal equilibrium. Study of the generalized Einstein coefficients and population dynamics of the two-level system in the thermal radiation field having large photon number fluctuation is kept as an open problem.  

\section*{Acknowledgement}
S. Biswas acknowledges partial financial support of the SERB, DST, Govt. of India under the EMEQ Scheme [No. EEQ/2019/000017]. Useful discussions with Prof. J. K. Bhattacharjee (IACS, Kolkata) are gratefully acknowledged.

\end{document}